\newcounter{myctr}
\def\myitem{\refstepcounter{myctr}\bibfont\noindent\ifnum\themyctr>9\else\phantom{0}\fi\hangindent17pt\themyctr.\enskip}

\documentclass{ws-ijqi}
\usepackage[dvipsnames]{xcolor}
\usepackage[super,sort,compress]{cite}
\usepackage{hyperref}
\usepackage{orcidlink}

\newcommand{\FA}[1]{\textcolor{black}{#1}}
\newcommand{\mga}[1]{\textcolor{black}{#1}}

\begin{document}

\catchline{}{}{}{}{}

\title{Statistical phase-space complexity of continuous-variable quantum channels}

\author{Siting Tang$^{1,2,3}$\,\orcidlink{0009-0001-9116-3889}, Francesco Albarelli$^{4}$\,\orcidlink{0000-0001-5775-168X}, Yue Zhang$^{1,2}$\,\orcidlink{0000-0003-4012-537X}, Shunlong Luo$^{1,2}$\,\orcidlink{0000-0003-4362-051X}, Matteo G. A. Paris$^{3,*}$\,\orcidlink{0000-0001-7523-7289}}

\address{$^{1}$State Key Laboratory of Mathematical Sciences, Academy of Mathematics and Systems Science, Chinese Academy of Sciences, Beijing 100190, China\\
$^{2}$School of Mathematical Sciences, University of Chinese Academy of Sciences,
Beijing 100049, China\\
$^{3}$Dipartimento di Fisica, Universit\`a  di Milano, I-20133 Milano, Italy\\
$^{4}$Universit\`a di Parma, Dipartimento di Scienze Matematiche, Fisiche e Informatiche, I-43124 Parma, Italy\\
$^{*}$matteo.paris@fisica.unimi.it}

\maketitle

\begin{history}
\received{\today}
\end{history}

\begin{abstract}
The statistical complexity of continuous-variable quantum states can be characterized with a quantifier defined in terms of information-theoretic quantities derived from the Husimi $Q$-function.
In this work, we utilize this complexity quantifier of quantum states to study the complexity of \FA{single-mode bosonic} quantum channels.
We define the complexity of quantum channels as the maximal amount of complexity they can generate from an initial state with the minimal complexity.
We illustrate this concept by evaluating the complexity of Gaussian channels and some examples of non-Gaussian channels.
\end{abstract}



\markboth{S. Tang, F. Albarelli, Y. Zhang, S. Luo and M. G. A. Paris}
{Statistical phase-space complexity of continuous-variable quantum channels}

\section{Introduction}	
Complexity is a ubiquitous and versatile concept in science, and various notions such as computational complexity \cite{Cook,Pap,Arora}, 
Kolmogorov complexity \cite{LiM,Shen}, circuit complexity \cite{Yao,Bhat}, communication complexity \cite{Buhrman,Brassard2003,Montina2012}, statistical complexity \cite{Bu2,Bu1}, etc, have been introduced to characterize and quantify different aspects of complexity in different contexts. Complexity of states and channels have been investigated from various perspectives \cite{Nielsen,Cubitt2012,Levene2018,Brenner2025,Ansu2024,Araiza2023,Nuradha2025,Li2025}. 

We are interested in complexity of quantum channels in this work, which can be characterized and quantified from many different perspectives. For instance, Montina \cite{Montina2012} quantified it using the minimal amount of classical communication, Levene {\it et al} \cite{Levene2018} considered the dimension of the Hilbert space, and Araiza {\it et al} \cite{Araiza2023} used a resource-dependent approach, to name but a few \cite{Brassard2003,Cubitt2012,Nuradha2025,Li2025}. In general, the complexity of quantum systems is a scenario-dependent concept and typically involves the tradeoff and interplay between classical and quantum features. Here we are going to consider it from a statistical point of view via the concept of Fisher information and Wehrl entropy.

Recently we introduced a quantifier of the  statistical phase-space complexity of a continuous-variable (CV) quantum state based on the Husimi $Q$-functions \cite{Husimi1940} $Q(\alpha |\rho) =\langle \alpha |\rho|\alpha\rangle$ of a single-mode continuous-variable (optical) state $\rho$. Here $|\alpha\rangle, \ \alpha \in \mathbb{C}$ are the Glauber coherent states. The complexity of the state $\rho$ is defined as \cite{Tang2025} 
\begin{equation}
    \label{eq:state_complexity}
    \mathcal{C} (\rho) = e^{S_\text{W} (\rho) -1} I(\rho),
\end{equation}
where
\begin{equation}
    S_\text{W} (\rho) = -\int_\mathbb{C} Q(\alpha |\rho) \ln Q(\alpha |\rho) \frac{\text{d}^2 \alpha}{\pi}
\end{equation}
is the Wehrl entropy \cite{Wehrl1978,Wehrl1979} of the state $\rho$, and
\begin{equation}
    I(\rho) = \frac{1}{4} \int_{\mathbb{C}}  
\frac{{\Vert {\nabla} Q(\alpha | \rho) \Vert^2}}{Q(\alpha | \rho)} \frac{\mathrm{d}^{2}\alpha}{\pi}
\end{equation}
is its Fisher information \cite{Fisher1925,Cramer1999,Stam1959} with respect to location parameters. Here $||\cdot ||$ is the usual Euclidean norm, and ${\rm d}^2\alpha ={\rm d}x{\rm d}y$ is the standard Lebesgue measure on $\mathbb{C}\sim \mathbb{R}^2$ with $\alpha =x+iy, x,y\in \mathbb{R}.$

\mga{The quantifier $\mathcal{C} (\rho)$  characterizes the statistical complexity of the CV state $\rho$ in terms of coherent states. 
While complexity is minimal for Gaussian mixture of coherent states (i.e., displaced thermal states), it increases for non-Gaussian mixtures, or as a result of operations that generate quantumness. This suggests that $\mathcal{C} (\rho)$ may have an operational meaning, given that quantum states are definitely resources for quantum information processing, and it has recently been shown that also non-Gaussian mixtures of Gaussian states can be advantageous for certain discrimination tasks \cite{gerry25}.}
%
%

In this work we would like to characterize the ability of a CV quantum channel to generate complexity, i.e., given a minimal-complex state, can a quantum channel help to increase the complexity of this state? This inspires us to define the complexity of a channel ${\cal E}$ as follows,
\begin{equation}
    \mathcal{C}({\cal E})  = \sup_{\rho_0: \ \mathcal{C}(\rho_0)=1} \mathcal{C} ({\cal E} (\rho_0)),
\end{equation}
where $\mathcal{C}(\rho)$ is the complexity of the state $\rho$ defined in Eq. (\ref{eq:state_complexity}). Note that in Refs.~[\citen{Dembo2002,Tang2025}] it was established that $\mathcal{C}(\rho) \geq 1$, with equality attained if and only if $\rho=D_{\xi} \nu_{\bar{n}} D_{\xi}^\dag$ is a displaced thermal state. Here $$D_{\xi} =e^{\xi a^\dag -\xi^* a},\qquad \xi\in\mathbb{C}$$ are the displacement operators, and 
\begin{equation}
\nu_{\bar{n}}=\frac{1}{1+\bar{n}} \sum_{k=0}^\infty \left(\frac{\bar{n}}{1+\bar{n}}\right)^k |k\rangle \langle k| \label{vv}
\end{equation}
is a thermal state with mean photon number $\bar{n}\geq0$. The above definition can be equivalently written as
\begin{equation}
    \mathcal{C}({\cal E}) = \sup_{\xi, \bar{n}} \mathcal{C}  ( {\cal E} (D_{\xi} \nu_{\bar{n}} D_{\xi}^\dag ) ).
\end{equation}

In the following Sections we are going to consider the Gaussian channel\FA{s} and some examples of non-Gaussian channel\FA{s} and study their ability to generate complexity.

\section{Gaussian channels}


\FA{We employ a physical parametrization of single-mode Gaussian channels, defining them as the dynamical maps arising from the following Lindblad master equation describing a diffusive Markovian dynamics~\cite{Walls1994,SerafiniBook}}
\begin{equation}
    \label{eq:channel_G}
    \frac{\text{d}\rho}{\text{d}t} = \frac{\Gamma}{2} \left\{ 
    (N+1)\mathcal{L}(a) +N\mathcal{L}(a^\dag ) -M^*\mathcal{D}(a) -M\mathcal{D}(a^\dag) \right\} \rho,
\end{equation}
where $$\mathcal{L}(X)\rho=2X\rho X^\dag -X^\dag X\rho -\rho X^\dag X,\quad \mathcal{D}(X)\rho = 2X\rho X -X^2\rho -\rho X^2$$ for any operator $X$, while $N \in \mathbb{R}$ and $M \in \mathbb{C}$ satisfy the constraint 
$|M|^2 \leq N(N+1).$

The channel can also be parameterized by its asymptotic state, which is a centered Gaussian state $S_{\eta} \nu_{\bar{n}} S_{\eta}^\dag$. Here $S_{\eta}=e^{\frac12(\eta a^{\dag 2} - \eta^* a^2)}$, $\eta=re^{i\theta}$  are the squeezing operators. Note that a thermal state $\nu_{\bar{n}}$ can be equivalently described by its purity $\mu = \text{tr} (\nu_{\bar{n}}^2)= \frac{1}{1+2\bar{n}}$. The parameterization of the above Gaussian channel in terms of its asymptotic state is \cite{Serafini2005,Ferraro2005}
\begin{eqnarray*}
    \mu_\infty &=& \frac{1}{\sqrt{(2N+1)^2-4|M|^2}}, \\
    \cosh(2r_\infty) &=& \sqrt{1 +\frac{4|M|^2}{(2N+1)^2-4|M|^2}}, \\
    \tan (\theta_\infty) &=& \tan (\text{arg} M).
\end{eqnarray*}

By solving Eq. (\ref{eq:channel_G}) with the initial state $\rho_0=D_{\xi} \nu_{\bar{n}} D_{\xi}^\dag$, one finds that the evolving state $\rho_t = {\cal E}_t (\rho_0 )$ maintains its Gaussian character. Additionally, we have that for any Gaussian state $\rho_{\text{g}} = D_{\xi} S_{\eta} \nu_{\bar{n}} S_{\eta}^\dag D_{\xi}^\dag$,  complexity can be explicitly expressed in terms of its purity $\mu$ and squeezing parameter $r$ as
\begin{equation*}
    \mathcal{C} (\rho_{\text{g}}) = \frac{1+\frac{1}{\mu}\cosh(2r)}{\sqrt{1+ \frac{2}{\mu}\cosh(2r)+ \frac{1}{\mu^2}}}.
\end{equation*}
Therefore for Gaussian channels, it holds that  
$$\mathcal{C} ({\cal E}) = \sup_{\xi, \bar{n}} \mathcal{C}  ( {\cal E} (D_{\xi} \nu_{\bar{n}} D_{\xi}^\dag ) )=\sup_{\bar n} \mathcal{C}  ({\cal E} (\nu_{\bar n}) ),$$ with $\nu_{\bar n}$ defined by Eq. (\ref{vv}). We only need the time evolution of the parameters $\mu$ and $r$, which has been solved in Ref.~\citen{Serafini2005},
\begin{eqnarray*}
    \mu(t) &=& \mu_0 \left[ \frac{\mu_0^2}{\mu_\infty^2} (1-e^{-\Gamma t})^2 + e^{-2\Gamma t} + 2\frac{\mu_0}{\mu_\infty} \cosh (2r_\infty) (1-e^{-\Gamma t})e^{-\Gamma t}
    \right]^{-\frac12}, \\
    \frac{\cosh (2r(t))}{\mu(t)} &=& \frac{1}{\mu_0} e^{-\Gamma t} + \frac{\cosh (2r_\infty)}{\mu_\infty} (1-e^{-\Gamma t}),
\end{eqnarray*}
where $\mu_0 =\frac{1}{1+2\bar{n}}$ is the purity of the initial state $\rho_0$.

We calculate the complexity of the Gaussian channel at time $t$,
\begin{eqnarray}
    \mathcal{C} ({\cal E}_t) &=& \sup_{\mu_0}\mathcal{C} (\rho_t) \nonumber\\
    &=& \sup_{\mu_0} \frac{1+\frac{1}{\mu(t)}\cosh(2r(t))}{\sqrt{1+ \frac{2}{\mu(t)}\cosh(2r(t))+ \frac{1}{\mu(t)^2}}} \nonumber\\
    &=& \sup_{\mu_0} \frac{1 +\frac{1}{\mu_0}e^{-\Gamma t} +(2N+1)(1-e^{-\Gamma t})}{\sqrt{\left[1 +\frac{1}{\mu_0}e^{-\Gamma t} +(2N+1)(1-e^{-\Gamma t})\right]^2 -4|M|^2(1-e^{-\Gamma t})^2}} \nonumber\\
    &=& \frac{1 +e^{-\Gamma t} +(2N+1)(1-e^{-\Gamma t})}{\sqrt{\left[1 +e^{-\Gamma t} +(2N+1)(1-e^{-\Gamma t})\right]^2 -4|M|^2(1-e^{-\Gamma t})^2}} \nonumber\\
    &=& \frac{1}{\sqrt{1-4\left(\frac {|M|}{{\rm coth} \left( \frac{\Gamma t}{2} \right) +2N+1}\right)^2 }}.
\end{eqnarray}
Note that in the second last line, the maximum is achieved at $\mu_0=1$, i.e., when the initial state $\rho_0 =|\alpha \rangle \langle \alpha |$ is a coherent state. We point out that for a non-squeezed bath ($M=0,$ or equivalently, $r_\infty=0$), the complexity stays constant,
$\mathcal{C}({\cal E}_t)=1, \forall \ t$, i.e., the bath never produces any complexity. On the other hand, for any fixed $N$, we note that $\mathcal{C} ({\cal E}_t)$ increases monotonically when $M$ increases, i.e., a Gaussian channel with a stronger degree of squeezing is able to produce more complexity. However, due to the constraint $|M|^2 \leq N(N+1)$, the complexity is upper bounded at any time $t$.
 
Let $t\to\infty$, we get
\begin{equation}
    \mathcal{C} ({\cal E}_\infty ) = \frac{1}{\sqrt{1-\left(\frac{|M|}{N+1}\right)^2}}.
\end{equation}
In this case, we have $\mathcal{C} ({\cal E}_\infty )\leq \sqrt{N+1}$. Equality is attained when $|M|^2=N(N+1)$, which is equivalent to $\mu_\infty=1$, i.e., the maximal complexity is produced when the asymptotic state is a pure squeezed state. This is consistent with the result in Ref.~\citen{Tang2025} that among all Gaussian states with a fixed energy, the most complex ones are the pure squeezed states, while the least complex ones are the displaced thermal states.

\section{Phase diffusion channels}

In this section we consider the phase diffusion channels \cite{Milburn1991}. Suppose the initial state is a displaced thermal state $\rho_0 = D_\xi \nu_{\bar{n}} D_\xi^\dag$ (without loss of generality, we assume $\xi\geq0$).
The Husimi $Q$-function for this state is a Gaussian distribution
\begin{equation}
Q(\alpha| \rho_0) = \langle \alpha | \rho_0 | \alpha \rangle = \langle \alpha | D_\xi \nu_{\bar{n}} D_\xi^\dag | \alpha \rangle =
\frac{1}{\bar{n} + 1} e^{ -\frac{|\alpha - \xi|^2}{\bar{n} + 1}}.
\end{equation}
Phase diffusion corresponds to applying random phase rotations $e^{-i\theta \hat{n}},$ where $\hat{n}=a^\dag a$. Let us model this using the von Mises distribution \cite{Risken1989} (the circular analogue of a Gaussian distribution)
\[
p(\theta) = \frac{e^{\kappa \cos \theta}}{2\pi I_0(\kappa)}, \qquad \theta \in [-\pi, \pi),
\]
where $\kappa \geq 0$ is the concentration parameter, and $I_0(\kappa)$ is the modified Bessel function of the first kind.
The phase-diffused state is
\begin{equation}
    \label{eq:def_phase_diffused_DTS}
    \rho_\kappa (\xi,\bar{n}) = \int_{-\pi}^\pi p(\theta) e^{-i\theta \hat{n}} \rho_0 e^{i\theta \hat{n}} \text{d}\theta.
\end{equation}
The Husimi $Q$-function of the evolved state is given by (we denote $Q(\alpha | \rho_\kappa (\xi,\bar{n}))=Q_\kappa(\alpha |\xi,\bar{n})$ for short)
\[
Q_\kappa(\alpha |\xi,\bar{n}) =\langle \alpha | \rho_\kappa (\xi,\bar{n}) | \alpha \rangle 
= \int_{-\pi}^{\pi} p(\theta) \langle \alpha | e^{-i\theta \hat{n}} \rho_0 e^{i\theta \hat{n}} | \alpha \rangle \text{d}\theta,
\]
Using  $e^{i\theta \hat{n}} | \alpha \rangle = | e^{i\theta} \alpha \rangle$, we get
\[
\langle \alpha | e^{-i\theta \hat{n}} \rho_0 e^{i\theta \hat{n}} | \alpha \rangle = \langle e^{i\theta} \alpha | \rho_0 | e^{i\theta} \alpha \rangle =  Q(e^{i\theta} \alpha |\rho_0),
\]
and therefore,
\begin{eqnarray*}
Q_\kappa(\alpha |\xi,\bar{n}) &=& \int_{-\pi}^{\pi} p(\theta) Q(e^{i\theta} \alpha |\rho_0) \text{d}\theta \\
&=& \int_{-\pi}^{\pi} \frac{e^{\kappa \cos \theta}}{2\pi I_0(\kappa)} \cdot \frac{1}{\bar{n} + 1} e^{ -\frac{|e^{i\theta} \alpha - \xi|^2}{\bar{n} + 1} } \text{d}\theta.
\end{eqnarray*}
Let us now express $\alpha$ in polar coordinate $\alpha = r e^{i\phi}$, so $e^{i\theta} \alpha = r e^{i(\phi + \theta)}$. Then
\[
|e^{i\theta} \alpha - \xi|^2 = |r e^{i(\phi + \theta)} - \xi|^2 = r^2 + \xi^2 - 2r\xi \cos(\phi + \theta),
\]
and thus
\[
Q_\kappa(r, \phi |\xi,\bar{n}) = \frac{1}{2\pi (\bar{n} + 1) I_0(\kappa)} e^{ -\frac{r^2 + \xi^2}{\bar{n} + 1} } \int_{-\pi}^{\pi} e^{\kappa \cos \theta} e^{ \frac{2r\xi \cos(\phi + \theta)}{\bar{n} + 1} } \text{d}\theta.
\]
If one defines the parameter
\begin{equation*}
\kappa' = \frac{2r\xi}{\bar{n} + 1},
\end{equation*}
then the integral becomes
\begin{equation*}
I = \int_{-\pi}^{\pi} e^{\kappa \cos \theta + \kappa' \cos(\phi + \theta)} \text{d}\theta.
\end{equation*}
Combining the exponents we get
\begin{eqnarray*}
    \kappa \cos \theta + \kappa' \cos(\phi + \theta) &=& \kappa \cos \theta + \kappa' (\cos\phi \cos\theta - \sin\phi \sin\theta)\\
    &=& (\kappa + \kappa' \cos\phi) \cos\theta + (-\kappa' \sin\phi) \sin\theta \\
    &=& R \cos(\theta - \Theta),
\end{eqnarray*}
where
\begin{eqnarray*}
    R &=& \sqrt{(\kappa + \kappa' \cos\phi)^2 + (\kappa' \sin\phi)^2} = \sqrt{\kappa^2 + \kappa'^2 + 2\kappa\kappa' \cos\phi},\\
    \Theta &=& \arctan\left( \frac{-\kappa' \sin\phi}{\kappa + \kappa' \cos\phi} \right).
\end{eqnarray*}
Thus,
\[
I = \int_{-\pi}^{\pi} e^{R \cos(\theta - \Theta)} \text{d}\theta = 2\pi I_0(R).
\]
Substituting back, we finally obtain the expression of the Husimi $Q$-function of a displaced thermal state undergoing phase diffusion described by a von Mises distribution,
\begin{eqnarray*}
    Q_\kappa(r, \phi |\xi,\bar{n}) 
    &=& \frac{1}{(\bar{n} + 1) I_0(\kappa)} e^{ -\frac{r^2 + \xi^2}{\bar{n} + 1} } I_0(R)\\
    &=& \frac{1}{(\bar{n} + 1) I_0(\kappa)} e^{ -\frac{r^2 + \xi^2}{\bar{n} + 1} } I_0 \left(\sqrt{\kappa^2 + \left( \frac{2r\xi}{\bar{n} + 1} \right)^2 + 2\kappa \left( \frac{2r\xi}{\bar{n} + 1} \right) \cos\phi} \right),
\end{eqnarray*}
or in complex coordinate,
\begin{equation}
    Q_\kappa(\alpha |\xi,\bar{n}) = \frac{1}{(\bar{n} + 1) I_0(\kappa)} e^{ -\frac{|\alpha|^2 + \xi^2}{\bar{n} + 1} } I_0 \left(\sqrt{\kappa^2 + \frac{4|\alpha|^2 \xi^2}{(\bar{n} + 1)^2}  + \frac{2\kappa\xi (\alpha+\alpha^*)}{\bar{n} + 1}} \right).
\end{equation}

Now we are going to maximize the complexity of the evolved state $\rho_\kappa (\xi,\bar{n})$ for fixed $\kappa$. Suppose the maximal complexity is attained at $(\xi_0,\bar{n}_0)$, i.e., $\sup_{\xi,\bar{n}} \mathcal{C} (\rho_\kappa (\xi,\bar{n})) = \mathcal{C} (\rho_\kappa (\xi_0,\bar{n}_0))$. Then consider another state $\rho_\kappa(\xi_0/\sqrt{\bar{n}_0+1},0)$, whose Husimi $Q$-function is
\begin{equation*}
    Q_\kappa \left(\alpha \Big| \frac{\xi_0}{\sqrt{\bar{n}_0+1}},0\right) = \frac{1}{I_0(\kappa)} e^{-|\alpha|^2-\frac{\xi_0^2}{\bar{n}_0+1}} I_0 \left( \sqrt{\kappa^2 + \frac{4|\alpha|^2\xi_0^2}{\bar{n}_0+1} + \frac{2\kappa\xi_0(\alpha+\alpha^*)}{\sqrt{\bar{n}_0+1}}} \right).
\end{equation*}
Let $\lambda=1/\sqrt{\bar{n}_0+1}$, and note that 
\begin{equation*}
    Q_\kappa (\alpha | \xi_0,\bar{n}_0) = \lambda^2 Q_\kappa \left(\lambda \alpha \Big| \frac{\xi_0}{\sqrt{\bar{n}_0+1}},0 \right).
\end{equation*}
So by the scaling invariance property \cite{Tang2025} of the complexity measure $\mathcal{C}(\cdot)$, we have
\begin{equation*}
    \mathcal{C} (\rho_\kappa (\xi_0,\bar{n}_0)) = \mathcal{C} \left(\rho_\kappa \left(\frac{\xi_0}{\sqrt{\bar{n}_0+1}},0\right)\right),
\end{equation*}
which means the maximum is also attained at $(\xi_0/\sqrt{\bar{n}_0+1},0)$. Therefore it suffices to consider $\sup_\xi \mathcal{C} (\rho_\kappa (\xi,0))$.

\begin{figure}
    \centering
    \includegraphics[width=0.48\linewidth]{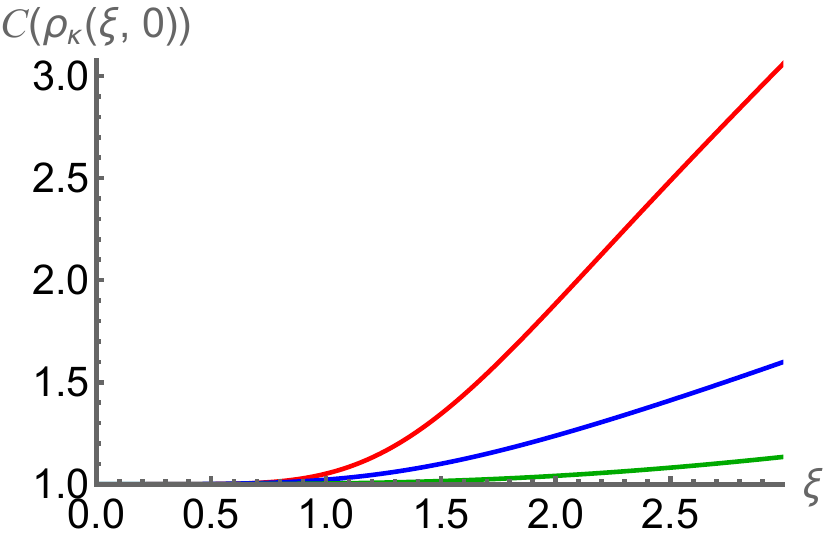}
    \includegraphics[width=0.48\linewidth]{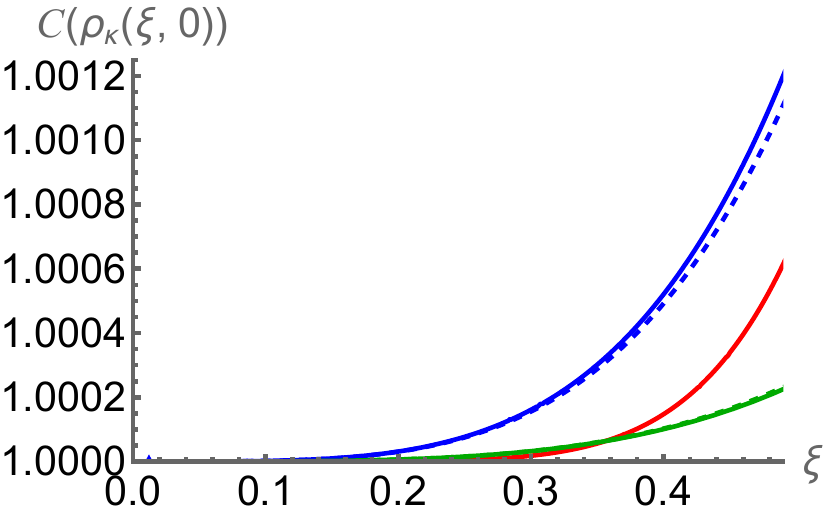}
    \caption{(a) The complexity of the phase-diffused state $\mathcal{C} (\rho_\kappa (\xi,0))$ as a function of $\xi$ with $\kappa=0.001,3,10$ (from top to bottom). (b) Blowup of (a) for small $\xi$, and the dashed lines are the fitted functions $1+\gamma_\kappa \xi^4$.}
    \label{fig:C_vs_xi}
\end{figure}
\begin{figure}
    \centering
    \includegraphics[width=0.5\linewidth]{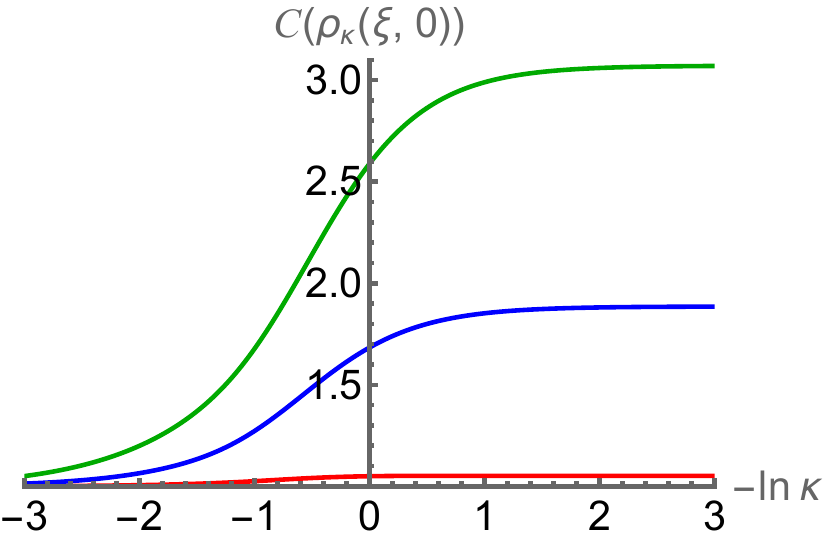}
    \caption{The complexity of the phase-diffused state $\mathcal{C} (\rho_\kappa (\xi,0))$ as a function of $\ln\frac 1\kappa =-\ln\kappa$ with $\xi=1,2,3$ (from bottom to top).}
    \label{fig:C_vs_kappa}
\end{figure}

In Fig. \ref{fig:C_vs_xi}(a) we plot the complexity $\mathcal{C} (\rho_\kappa (\xi,0))$ as a function of $\xi$ for different values of $\kappa$. Note that $\kappa\geq0$ is the concentration parameter, where a large value of $\kappa$ corresponds to a small degree of phase diffusion. As long as $\kappa<\infty$, the channel is non-Gaussian. We see that the complexity increases in the displacement parameter $\xi$, and as $\xi \to \infty$, the complexity is unbounded. Therefore we find that a small non-Gaussianity is enough to give the channel the ability to produce an unbounded complexity, i.e., we have
\begin{equation}
    \mathcal{C} ({\cal E}_\kappa ) = \infty.
\end{equation}
However, in order to generate a large complexity, we need an initial state $\rho_0$ with some energy, i.e., $\xi$ cannot be too small. From Fig. \ref{fig:C_vs_xi}(a) we see that when $\xi \gg 1$, the growth is almost linear in $\xi$, and a smaller $\kappa$ (i.e., a larger non-Gaussianity) gives a faster growth. We illustrate this by plotting $\mathcal{C} (\rho_\kappa (\xi,0))$ in terms of $\kappa$ in Fig \ref{fig:C_vs_kappa}. We see that the complexity is increasing in $\ln\frac 1\kappa$, and thus decreasing in $\kappa$. If we are given a fixed initial state $\rho_0=|\xi\rangle\langle\xi|$ with $\xi\gg1$, then the phase diffusion channel ${\cal E}_\kappa$ produces larger complexity as $\kappa\to0$.

On the other hand, the behavior of the complexity for $\xi \ll 1$ is enlarged in Fig. \ref{fig:C_vs_xi}(b). In fact, if we expand the complexity $\mathcal{C} (\rho_\kappa (\xi,0))$ in Taylor series, we will find
\begin{equation*}
    \mathcal{C} (\rho_\kappa (\xi,0)) = 1+\gamma_\kappa \xi^4 + o(\xi^4),
\end{equation*}
where the coefficient of the fourth degree term depends on $\kappa$, and reads
\begin{equation*}
    \gamma_\kappa=\frac{1}{2} \left( \frac{2I_0(\kappa)I_1(\kappa)+\kappa I_1(\kappa)^2}{\kappa I_0(\kappa)^2}-1 \right)^2.
\end{equation*}
Note that $\gamma_\kappa$ first increases in $\kappa$ and then decreases. This explains the behavior in Fig. \ref{fig:C_vs_xi}(b) that $\kappa=3$ (blue line) gives the fastest growth for small $\xi$. So when $\xi\ll1$, the optimal value of $\kappa$ that enables the channel to produce the largest complexity is nontrivial.

\section{Photon addition and photon subtraction}

In this section we discuss the effects of \emph{successful} photon addition and photon subtraction. Experimentally they are usually realized by preparing an ancillary mode in the vacuum state and using a weak parametric amplifier (for addition) or a weakly reflecting beam splitter (for subtraction), with the detection of a photon in the ancillary mode heralding a successful addition or subtraction \cite{Walls1994,Agarwal2013}. We realize that these are only probabilistic events, however the overall channels are Gaussian \cite{Barnett2002}, so given an initial state as the displaced thermal the output states are still displaced thermal states, thus the overall channels do not create any complexity. Therefore we put our interest in the ideal sub-channels corresponding to successful photon addition or subtraction, and they are apparently non-Gaussian.

\begin{figure}
    \centering
    \includegraphics[width=0.48\linewidth]{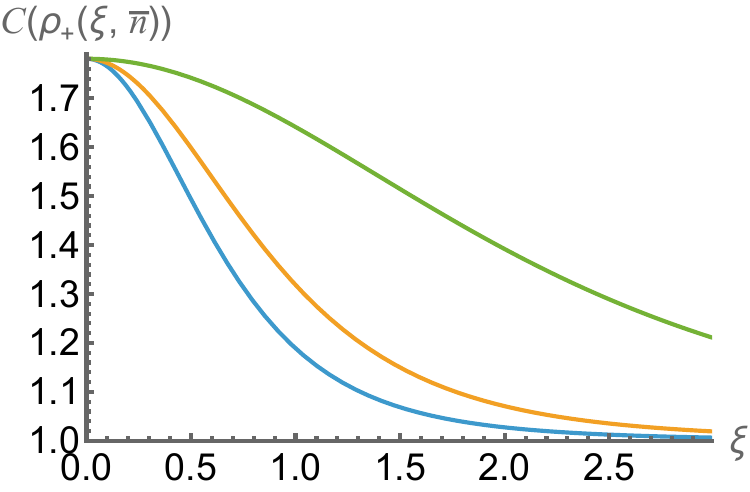}
    \includegraphics[width=0.48\linewidth]{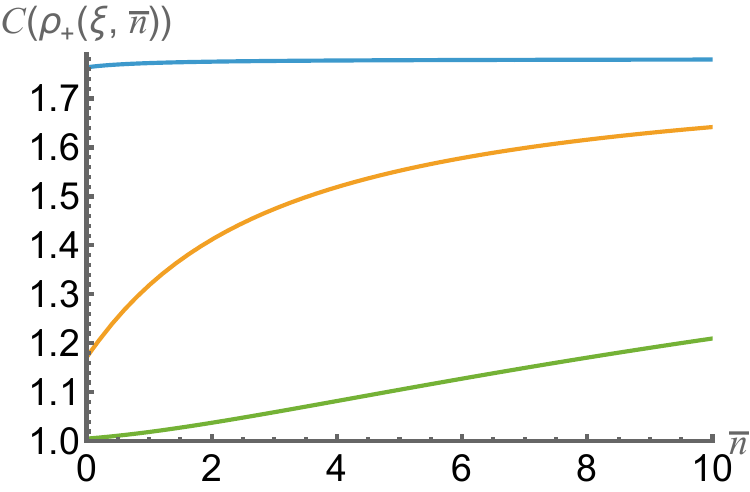}
    \caption{(a) The complexity of the photon-added displaced thermal state $\mathcal{C} (\rho_+ (\xi,\bar{n}))$ as a function of $\xi$ with $\bar{n}=0.1,1,10$ (from bottom to top). (b) The complexity of the photon-added displaced thermal state $\mathcal{C} (\rho_+ (\xi,\bar{n}))$ as a function of $\bar{n}$ with $\xi=0.1,1,3$ (from top to bottom).}
    \label{fig:PA}
\end{figure}

As usual we start with the displaced thermal state $\rho_0 = D_\xi \nu_{\bar{n}} D_\xi^\dag$, then photon addition leads to the state
\begin{equation}
   {\cal E}_+(\rho_0 ) = \frac{1}{1+\xi^2+\bar{n}} \, a^\dag  D_\xi \nu_{\bar{n}} D_\xi^\dag a =:\rho_+(\xi,\bar{n}) .
\end{equation}
Its $Q$-function can be easily obtained as
\begin{equation}
    Q_+(\alpha|\xi,\bar{n}) = \langle \alpha |\rho_+ (\xi,\bar{n}) |\alpha\rangle
    = \frac{|\alpha|^2}{(1+\xi^2+\bar{n})(1+\bar{n})} \, e^{-\frac{|\alpha-\xi|^2}{1+\bar{n}}}.
\end{equation}
With the expression of the Husimi $Q$-function, we can numerically calculate the complexity of the photon-added displaced thermal state. In Fig. \ref{fig:PA} we plot $\mathcal{C} (\rho_+ (\xi,\bar{n}))$ as a function of $\xi$ and $\bar{n},$ respectively. We observe that the complexity decreases in $\xi$ and increases in $\bar{n}$. Therefore the maximal complexity is achieved at $\xi=0$, giving us a photon-added thermal state $\nu_+ = \rho_+(0,\bar{n}) = \frac{1}{1+\bar{n}} a^\dag \nu_{\bar{n}} a$, whose complexity has been derived in Ref.~\citen{Tang2025} and is independent of the thermal photon number $\bar{n}$. Thus,
\begin{equation}
    \mathcal{C} ({\cal E}_+) = \sup_{\xi,\bar{n}} \mathcal{C} (\rho_+(\xi,\bar{n})) = \mathcal{C} (\nu_+) = e^\gamma,
\end{equation}
where $\gamma$ is the Euler constant.

\begin{figure}
    \centering
    \includegraphics[width=0.48\linewidth]{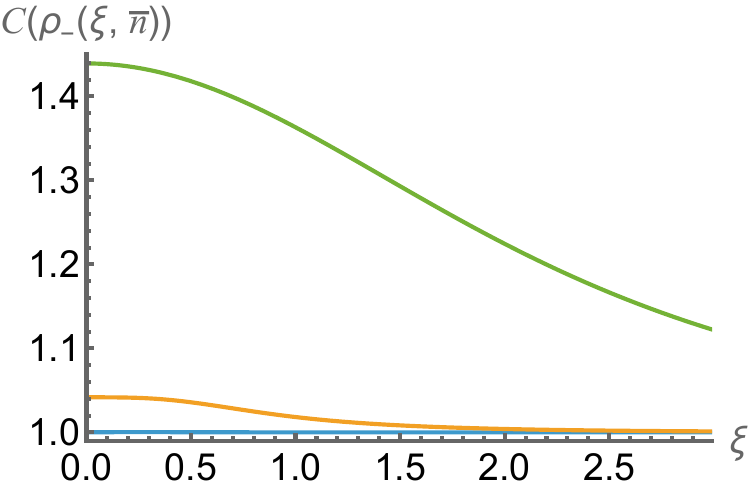}
    \includegraphics[width=0.48\linewidth]{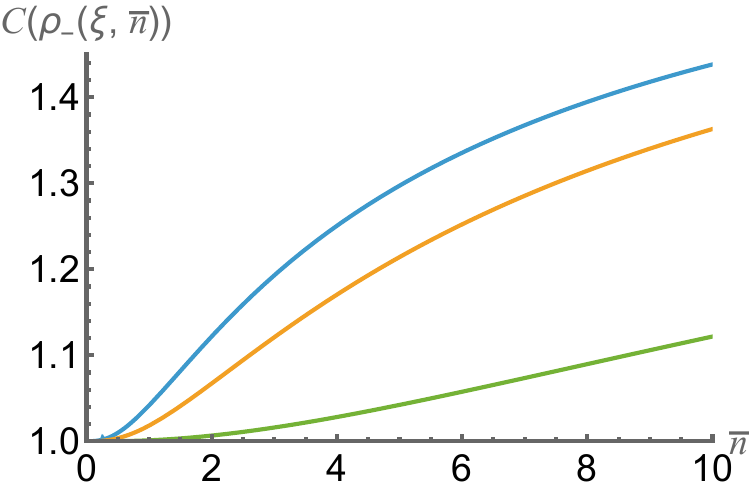}
    \caption{(a) The complexity of the photon-subtracted displaced thermal state $\mathcal{C} (\rho_- (\xi,\bar{n}))$ as a function of $\xi$ with $\bar{n}=0.1,1,10$ (from bottom to top). (b) The complexity of the photon-subtracted displaced thermal state $\mathcal{C} (\rho_- (\xi,\bar{n}))$ as a function of $\bar{n}$ with $\xi=0.1,1,3$ (from top to bottom).}
    \label{fig:PS}
\end{figure}

Next we consider applying photon subtraction to the displaced thermal state
\begin{equation}
   {\cal E}_-(\rho_0) = \frac{1}{\xi^2+\bar{n}}\, a D_\xi \nu_{\bar{n}} D_\xi^\dag a^\dag =: \rho_-(\xi,\bar{n}).
\end{equation}
The Husimi $Q$-function is given by
\begin{eqnarray*}
    Q_-(\alpha|\xi,\bar{n}) &=& \langle \alpha |\rho_- (\xi,\bar{n}) |\alpha\rangle \\
    &=& \frac{1}{\xi^2+\bar{n}} \sum_{k=0}^\infty \frac{1}{1+\bar{n}} \left(\frac{\bar{n}}{1+\bar{n}}\right)^k |\langle \alpha |a D_\xi |k\rangle |^2 \\
    &=& \frac{1}{\xi^2+\bar{n}} \sum_{k=0}^\infty \frac{1}{1+\bar{n}} \left(\frac{\bar{n}}{1+\bar{n}}\right)^k |\langle \alpha | D_\xi (a+\xi) |k\rangle |^2 \\
    &=& \frac{1}{\xi^2+\bar{n}} \sum_{k=0}^\infty \frac{1}{1+\bar{n}} \left(\frac{\bar{n}}{1+\bar{n}}\right)^k \left|\langle \alpha-\xi | \left(\sqrt{k} |k-1\rangle +\xi|k\rangle \right) \right|^2 .
\end{eqnarray*}
Denote $\beta=\alpha-\xi$, and note that complexity is invariant under phase-space displacements in the sense that ${\cal C} (D_\xi^\dag\rho D_\xi)={\cal C}(\rho)$, we have
\begin{eqnarray}
    Q_-(\beta|\xi,\bar{n}) &=& \frac{1}{\xi^2+\bar{n}} \sum_{k=0}^\infty \frac{1}{1+\bar{n}} \left(\frac{\bar{n}}{1+\bar{n}}\right)^k e^{-|\beta|^2} \frac{|\beta|^{2(k-1)}}{(k-1)!} \left( k+\frac{\xi^2|\beta|^2}{k} +\xi(\beta+\beta^*) \right) \nonumber\\
    &=& \frac{1}{\xi^2+\bar{n}}\, e^{-\frac{|\beta|^2}{1+\bar{n}}} \left( \frac{\bar{n}^2|\beta|^2}{(1+\bar{n})^3} + \frac{\bar{n}(1+\xi(\beta+\beta^*))}{(1+\bar{n})^2} + \frac{\xi^2}{1+\bar{n}}\right).
\end{eqnarray}
The complexity is evaluated numerically, and its properties are illustrated in Fig. \ref{fig:PS}. As seen in previous examples, the complexity decreases with $\xi$ and increases with $\bar{n}$. Hence in order to find the maximal complexity generated by photon subtraction, the parameters of the initial state should be set at $\xi=0$ and let $\bar{n} \to \infty$.

If we examine again the Husimi $Q$-function of the photon-subtracted thermal state $\nu_- :=\rho_-(0,\bar{n}) = \frac{1}{\bar{n}} a  \nu_{\bar{n}} a^\dag$, we will find
\begin{eqnarray*}
    Q(\beta|\nu_-) =Q_-(\beta|0,\bar{n}) &=& \frac{1}{\bar{n}} e^{-\frac{|\beta|^2}{1+\bar{n}}} \left( \frac{\bar{n}^2|\beta|^2}{(1+\bar{n})^3} + \frac{\bar{n}}{(1+\bar{n})^2} \right) \\
    &=& \frac{\bar{n}}{1+\bar{n}} \cdot \frac{1}{(1+\bar{n})^2} e^{-\frac{|\beta|^2}{1+\bar{n}}} |\beta|^2 + \frac{1}{1+\bar{n}} \cdot \frac{1}{1+\bar{n}} e^{-\frac{|\beta|^2}{1+\bar{n}}} \\
    &=& \frac{\bar{n}}{1+\bar{n}} Q (\beta | \nu_+) +\frac{1}{1+\bar{n}}  Q(\beta|\nu_{\bar{n}}),
\end{eqnarray*}
which is a mixture of the Husimi $Q$-functions of the photon-added thermal state $\nu_+$ and the thermal state $\nu_{\bar{n}}$. Since the Husimi $Q$-functions are faithful representations of quantum states, we may conclude that the photon-subtracted thermal state is a mixture as well
\begin{equation*}
    \nu_- = \frac{\bar{n}}{1+\bar{n}} \nu_+ + \frac{1}{1+\bar{n}} \nu_{\bar{n}}.
\end{equation*}
As $\bar{n} \to\infty$, we will have $\nu_- \to \nu_+$, whose complexity is $e^\gamma$. Therefore, the complexity generated by photon subtraction is also
\begin{equation}
    \mathcal{C} ({\cal E}_-) = \sup_{\xi,\bar{n}} \mathcal{C} (\rho_-(\xi,\bar{n})) = e^\gamma.
\end{equation}

\section{Conclusions}
In this work, we have advanced the quantification of quantum complexity from states to channels, building upon our previous framework that defines the complexity of a continuous-variable quantum state via its Husimi $Q$-function, 
Wehrl entropy, and Fisher information. The central question we addressed is: To what extent can a quantum channel generate complexity when acting on a minimal-complex state, i.e.,  a displaced thermal state?
We have introduced a natural definition for the complexity of a channel as the supremum of the output state 
complexity over all such minimal-complex input states. This allows us to probe the inherent {\em complexity-generating power} of a quantum channel.

For single-mode Gaussian channels\FA{, obtained as solutions of a diffusive open dynamics}, we have derived a closed-form, time-dependent expression for the channel 
complexity.
We have demonstrated that a non-squeezed bath cannot generate any complexity at any time. 
The maximum achievable complexity is bounded and is attained when the asymptotic state of the channel 
is a pure squeezed state, consistent with the known fact that pure squeezed states are the most complex 
among Gaussian states with fixed energy.

Phase diffusion (a non-Gaussian channel), as  modeled via the von Mises distribution, possesses an unbounded 
capacity to generate complexity. The complexity grows with the initial displacement of the input coherent 
state. Interestingly, for small displacements ($\xi \ll 1$), the growth rate of complexity with respect 
to $\xi$ is non-monotonic with the diffusion strength, indicating a non-trivial optimal regime for 
complexity generation at low energies.

Concerning photon addition and subtraction, we found that for the ideal, heralded sub-channels corresponding to successful photon addition and subtraction, the generated complexity is bounded and identical for both 
operations. This maximum is achieved for input thermal states (zero displacement) in the limit of high 
thermal photon number for subtraction.

Our results reveal a fundamental distinction: while Gaussian channels are limited in their complexity 
generation, the introduction of even simple forms of non-Gaussianity, such as phase diffusion, can 
lead to an unbounded complexity. This underscores the role of non-Gaussian operations as a critical 
resource for achieving high complexity in quantum systems, which is often linked to some form of operational 
advantage. Our results pave indeed the way to investigate the direct operational implications of channel 
complexity in quantum metrology, communication, and computation. 

The differences and relations between various notions of complexity are worth further investigation.

\section*{Acknowledgments}
This work has been done under the auspices of GNFM-INdAM and has been partially supported by MUR and EU through the projects PRIN22 RISQUE (CUP G53D23001110006), PRIN22-PNRR QWEST (CUP G53D23006270001), NQSTI-Spoke1-BaC QBETTER (CUP G43C22005120007), NQSTI-Spoke2-BaC QMORE (CUP J13C22000680006), the National Natural Science Foundation of China, Grant Nos. 12426671, 12401609, 12341103, and the Youth Innovation Promotion Association of CAS, Grant No. 2023004.

\end{document}